\documentclass[prl,twocolumn,nosuperscriptaddress]{revtex4}
\usepackage{graphicx}% Include figure files
\usepackage{dcolumn}% Align table columns on decimal point
\usepackage{amsmath,amsthm,amssymb}

\begin{document}
\title{Instability of Flat Space Enclosed in a Cavity}

\author{Maciej Maliborski} \affiliation{M. Smoluchowski Institute of
  Physics, Jagiellonian University, 30-059 Krak\'ow, Poland}

\date{\today}
\begin{abstract}
  We consider a spherically symmetric self-gravitating massless scalar
  field enclosed inside a timelike worldtube $\mathbb{R}\times S^3$
  with a perfectly reflecting wall. Numerical evidence is given that
  arbitrarily small generic initial data evolve into a black hole.
\end{abstract}

% \pacs{Valid PACS appear here}% PACS, the Physics and Astronomy
% Classification Scheme.  \keywords{Suggested
% keywords}%Use showkeys class option if keyword
% display desired
\maketitle

Bizo\'n and Rostworowski \cite{br} recently gave evidence that any
generic small perturbation of anti--de Sitter (AdS) spacetime evolves
into a black hole after a time $\mathcal{O}(\varepsilon^{-2})$, where
$\varepsilon$ measures the size of the perturbation. They argued that
this instability is a combined result of the following two
effects. First, the timelike boundary at spatial infinity of AdS acts
as a mirror at which the waves propagating outwards bounce back and
return to the bulk. Second, nonlinear resonant interactions of
reflected waves tend to shift their energy into increasingly high
frequencies, or equivalently increasingly fine spatial scales, which
eventually leads to the formation of a horizon. There arises a natural
question whether this kind of turbulent dynamics is specific to
asymptotically AdS spacetimes, or it can also occur in other
situations when dissipation of energy by dispersion is
absent. Motivated by this question, in this paper we consider small
perturbations of a portion of Minkowski spacetimes enclosed inside a
timelike worldtube $\mathbb{R}\times S^3$. Admittedly, this problem is
somewhat artificial geometrically, yet we think that it sheds some new
light on the results of \cite{br}.

We restrict ourselves to spherical symmetry and choose a minimally
coupled massless scalar field $\phi(t,r)$ as matter source. It is
convenient to parametrize the general spherically symmetric metric as
follows
\begin{equation}
  ds^2 = -\frac{A}{N^{2}} dt^2 + A^{-1} dr^2 + r^2 d\Omega^2\,,
\end{equation}
where $r$ is the areal radial coordinate, $d\Omega^2$ is the round
metric on the unit two-sphere, and the functions $A$ and $N$ depend on
$t$ and $r$. We introduce auxiliary variables $\Phi:= \phi'$ and
$\Pi:= A^{-1} N \dot\phi$ (hereafter primes and dots denote
$\partial_{r}$ and $\partial_{t}$, respectively) and write the wave
equation $\nabla^{\alpha} \nabla_{\alpha} \phi=0$ in the first order
form
\begin{equation}
  \label{eq:wave}
  \dot\Phi = \left(\frac{A\Pi}{N}\right)'\,, \quad \dot\Pi =
  \frac{1}{r^2}\left( r^2 \frac{A\Phi}{N} \right)'\,.
\end{equation}
Einstein's equations $G_{\mu\nu}= 8\pi G \left[\partial_{\mu} \phi
  \,\partial_{\nu} \phi - \frac{1}{2} g_{\mu\nu} (\partial
  \phi)^2\right]$ take a concise form when expressed in terms of the
mass function $m:=\frac{1}{2} r (1-A)$ (in units where $4\pi G=1$)
\begin{eqnarray}
  \label{eq:dprime}
  \frac{N'}{N} &=&-r \left( \Phi^2 + \Pi^2 \right)\,,\\
  \label{eq:mprime}
  m'&=&\frac{1}{2} r^2 A \left( \Phi^2 + \Pi^2 \right)\,,\\
  \label{eq:mdot}
  \dot m &=& r^2 \frac{A}{N} \Phi \Pi \,.
\end{eqnarray}
To ensure regularity at $r=0$ we require that
\begin{align}
  \Phi(t,r)=\mathcal{O}\left(r\right), \quad
  \Pi(t,r)=p_{0}(t)+\mathcal{O}\left(r^2\right), \\ m(t,r) =
  \mathcal{O}\left(r^3\right),\quad
  N(t,r)=1+\mathcal{O}\left(r^2\right)\,,
\end{align}
with expansion coefficients uniquely determined by a free function
$p_{0}(t)$. We set $N(t,0)=1$ so that $t$ is the proper time at the
center.

In the asymptotically flat situation the above system has been
extensively studied in the past leading to important insights about
the dynamics of gravitational collapse. In particular, Christodoulou
showed that for small initial data the fields disperse to infinity
\cite{ch1}, while for large initial data black holes are formed
\cite{ch2}. The borderline between these two generic outcomes of
evolution was explored numerically by Choptuik leading to the
discovery of critical phenomena at the threshold of black hole
formation \cite{ch}. The aim of this note is to see how these findings
are affected by placing a reflecting mirror at some finite radius
$r=R$. In other words, instead of asymptotically flat boundary
conditions, we consider the interior Dirichlet problem inside a ball
of radius $R$ with the boundary condition $\Pi(t,R)=0$.  The Dirichlet
boundary condition and the requirement of smoothness imply that the
coefficients of the power series expansions at $r=R$ (here
$\rho=1-r/R$)
\begin{eqnarray}\label{series}
  \phi(t,r)&=&\sum_{k\geq 1} \phi_k(t) \rho^k, \\
  N(t,r)&=&\sum_{k\geq 0} N_k(t) \rho^k,\quad
  m(t,r)=\sum_{k\geq 0} m_k(t) \rho^k \nonumber
\end{eqnarray}
are determined recursively by $\phi_1(t)$, $N_0(t)$, and $m_0(t)$.
For example, at the lowest order we have:
\begin{equation}
  \label{compatibility}
  2 \phi_2(t) - \left( 1 + \frac{R}{R-2m_0(t)} \right) \phi_1(t)=0\,.
\end{equation}
Taken at $t = 0$, the expansions \eqref{series} express the
compatibility conditions between initial and boundary values. It
follows immediately from equation \eqref{eq:mdot} that
$m_0(t)=M=\mathrm{const}$, where $M$ is the total energy. From
\eqref{eq:mprime} the total energy can be expressed as the volume
integral
\begin{equation}
  \label{eq:tm}
  M = \frac{1}{2} \int_0^R A  \left( \Phi^2 + \Pi^2 \right) r^2 dr\,.
\end{equation}

We solved the initial-boundary value problem for the system
\eqref{eq:wave}-\eqref{eq:mdot} numerically using the method of lines
with a fourth-order spatial finite-difference discretization
scheme. Time integration of evolution equations was performed with use
of an adaptive, explicit Runge-Kutta-Dormand-Prince algorithm of order
5(4). The metric functions were updated by solving the slicing
condition \eqref{eq:dprime} and the Hamiltonian constraint
\eqref{eq:mprime}. The degree to which the constraint \eqref{eq:mdot}
is preserved and the mass \eqref{eq:tm} is conserved depending on the
spatial resolution was used to verify the fourth order convergence of
the numerical scheme. Let us point out that this fully constrained
evolution scheme is very efficient computationally and easy to make
parallel because the update of metric functions reduces to simple
integrations. Namely, from \eqref{eq:dprime} we have
\begin{equation}
  \log N(t,r)=-\int_0^r s\,\left[ \Phi(t,s)^2 + \Pi(t,s)^2 \right] ds\,,
\end{equation}
and from the combination of \eqref{eq:dprime} and \eqref{eq:mprime} we
get
\begin{equation}
  A(t,r)=\frac{N(t,r)}{r} \int_0^r \frac{ds}{N(t,s)}\,.
\end{equation}

\begin{figure}[h]
  \centering
  \includegraphics[width=0.99\columnwidth]{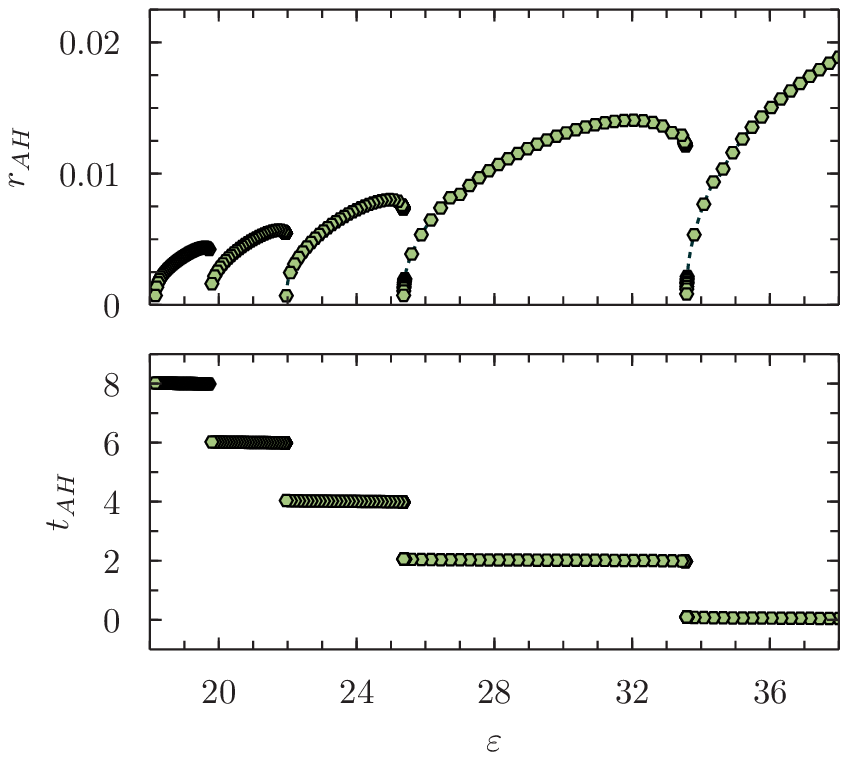}
  \caption{Apparent horizon radius $r_{AH}$ (top) and corresponding
    formation time $t_{AH}$ (bottom) as a function of the amplitude of
    initial data \eqref{eq:idg}. At critical points
    $\lim_{\varepsilon\rightarrow\varepsilon_{n}^+}
    r_{AH}(\varepsilon)=0$, while the horizon formation time exhibits
    jumps of size
    $t_{AH}(\varepsilon_{n+1})-t_{AH}(\varepsilon_{n})\approx 2$ (time
    in which the pulse traverses the cavity back and forth).}
  \label{fig:teeth}
\end{figure}

\begin{figure}[h]
  \centering
  \includegraphics[width=0.99\columnwidth]{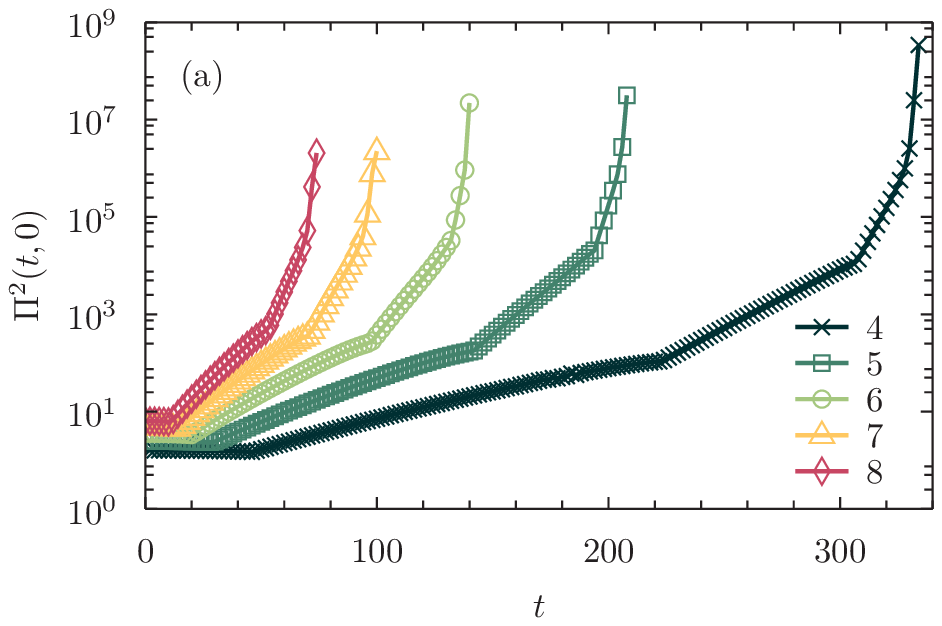}
  \vspace{1ex}

  \centering
  \includegraphics[width=0.99\columnwidth]{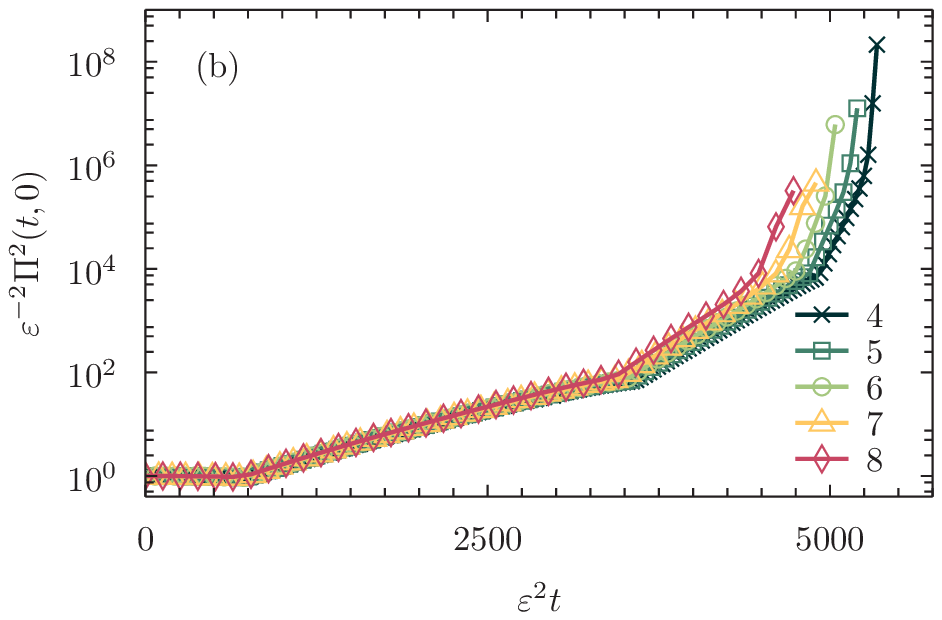}
  \caption{(a) The function $\Pi^{2}(t,0)$ for solutions with initial
    data \eqref{eq:idg} for several moderately small amplitudes. For
    clarity of the plot only the envelopes of rapid oscillations are
    depicted. (b) The curves from the plot (a) scaled according to
    $\varepsilon^{-2}\Pi^{2}(\varepsilon^{2} t,0)$. Plotted curves are
    labelled by the value of initial data amplitude $\varepsilon$.}
  \label{fig:ricci}
\end{figure}

Numerical results presented below were generated from Gaussian-type
initial data of the form (without loss of generality we set $R=1$)
\begin{equation}
  \label{eq:idg}
  \Phi(0,r)=0,\quad \Pi(0,r)=\varepsilon \exp\left( -32
    \tan^{2}\frac{\pi}{2} r \right)\,.
\end{equation}
These initial data vanish exponentially as $r\rightarrow 1$ so
compatibility conditions are not an issue.  The results are very
similar to those of \cite{br}, as can be seen by comparing
Figs.~\ref{fig:teeth} and \ref{fig:ricci} with the analogues figures
in \cite{br}.  For large amplitudes the evolution is not affected by
the mirror; the wave packet rapidly collapses, forming an apparent
horizon at a point where the metric function $A(t,r)$ goes to
zero. However, a wave packet which is marginally too weak to form a
horizon on the first implosion, does so on the second implosion after
being reflected back by the mirror.  As in the AdS case, this leads to
a sequence of critical amplitudes $\varepsilon_n$ for which the
solutions, after making $n$ bounces, asymptote Choptuik's critical
solution (see Fig.~\ref{fig:teeth}). To track the steepening of the
wave packet for very small amplitudes, we follow \cite{br} and monitor
the Ricci scalar at the center $R(t,0)=-2\Pi(t,0)^{2}$. This function
oscillates with approximate period $2$. Initially, the amplitude stays
almost constant but after some time it begins to grow exponentially
and eventually a horizon forms [see Fig.~\ref{fig:ricci}(a)]. As shown
in Fig.~\ref{fig:ricci}(b), the time of onset of exponential growth
$T$ scales with the amplitude of initial data as $T\sim
\varepsilon^{-2}$, which indicates that arbitrarily small
perturbations (for which it is impossible numerically to track the
formation of a horizon) eventually start growing.

In \cite{br} the numerical results were corroborated by a nonlinear
perturbation analysis which demonstrated that the instability of AdS
is caused by the resonant transfer of energy from low to high
frequencies. For the problem at hand, the evolution of linearized
perturbations is governed by the free radial wave equation $\ddot \phi
= r^{-2}(r^2 \phi')'$; hence the eigenfrequencies and eigenmodes are
\begin{equation}
  \omega^{2}_{j}=\left(\frac{j\pi}{R}\right)^{2}\,, \quad
  e_{j}(r)=\sqrt{\frac{2}{R}}\,\frac{\sin \omega_{j} r}{r}\,,\quad
  j\in\mathbb{N}\,.
  \label{eigen}
\end{equation}
As in AdS, the spectrum is fully resonant (that is, the frequencies
$\omega_{j}$ are equidistant), so the entire perturbation analysis of
\cite{br} can be formally repeated in our case.  We say 'formally'
because, in contrast to the AdS case, the eigenmodes \eqref{eigen}
violate the compatibility conditions at $r=R$ and therefore they
cannot be taken as smooth initial data.

\begin{figure}[!h]
  \centering
  \includegraphics[width=0.99\columnwidth]{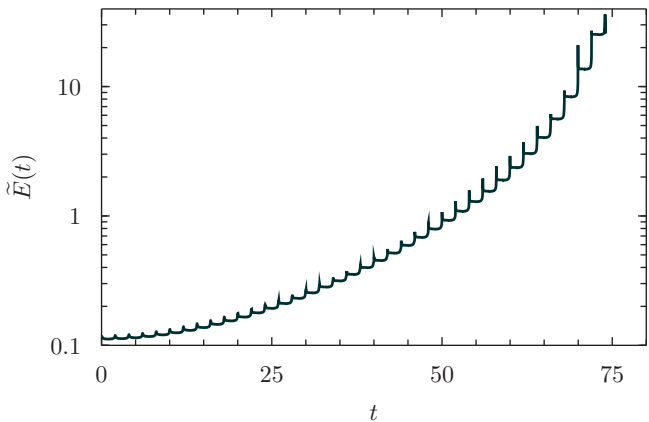}
  \caption{Plot of the weighted energy norm $\widetilde E(t)$ for the
    solution with initial amplitude $\varepsilon=8$.  The steep bursts
    of growth occur when the pulse implodes through the center.}
  \label{fig:sobolevnorm}
\end{figure}

The transfer of energy to higher modes (which is equivalent to the
concentration of energy on smaller scales) can be quantified by
monitoring the energy contained in the linear modes
$E_{j}=\Pi_{j}^{2}+\omega_{j}^{-2}\Phi_{j}^{2}$, where
$\Phi_{j}=\left(A^{1/2}\Phi,e'_{j}\right)$ and
$\Pi_{j}=\left(A^{1/2}\Pi,e_{j}\right)$, with the inner product
defined as $(f,g):=\int_{0}^{R}f(r)g(r)r^{2}dr$. Then, the total
energy can be expressed as the Parseval sum
$M=\sum_{j=1}^{\infty}E_j(t)$. The evidence for the energy transfer is
shown in Fig.~\ref{fig:sobolevnorm} which depicts a Sobolev-type
weighted energy norm \linebreak $\widetilde
E(t)=\sum_{j=1}^{\infty}j^{2}E_{j}(t)$.  The growth of $\widetilde
E(t)$ in time means that the distribution of energy shifts from low to
high frequencies.  The characteristic staircase shape of $\widetilde
E(t)$ indicates that the energy transfer occurs mainly during the
subsequent implosions through the center.  This observation leads to
the conclusion that the only role of the mirror is to reflect the
pulse so that it can be focused during the next implosion.

\begin{figure}[!h]
  \centering
  \includegraphics[width=0.99\columnwidth]{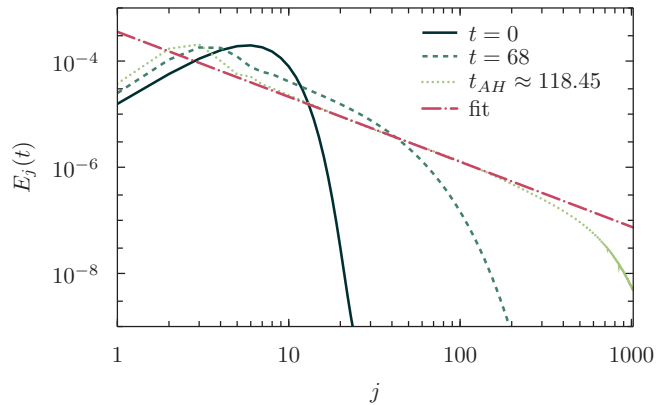}
  \caption{Energy spectra at three moments of time (initial,
    intermediate, and just before collapse) for the solution with
    initial amplitude $\varepsilon=6$.  The fit of the power law
    $E_{j}\sim j^{-\alpha}$ performed on the interval $j\in[16,128]$
    gives the slope $\alpha\approx 1.2$.}
  \label{fig:spectra}
\end{figure}

Another aspect of the turbulent cascade is shown in
Fig.~\ref{fig:spectra} which depicts the spectrum of energy (that is,
the distribution of the total energy over the linear modes) for the
solution with initial data \eqref{eq:idg} and $\varepsilon=6$.
Initially, the energy is concentrated in low modes; the exponential
decay of the spectrum expresses the smoothness of initial data. During
the evolution the range of excited modes increases and the spectrum
becomes broader. Just before horizon formation an intermediate range
of the spectrum exhibits the power-law scaling $E_{j}\sim j^{-\alpha}$
with exponent $\alpha = 1.2\pm0.1\,$. Energy spectra in evolutions of
different families of small initial data exhibit the same slope (up to
a numerical error) which indicates that the exponent $\alpha$ is
universal. We note that the power-law spectrum with a similar exponent
was also observed in the AdS case \cite{br2}. As pointed out in
\cite{br}, the black hole formation provides a cut-off for the
turbulent energy cascade for solutions of Einstein's equations (in
analogy to viscosity in the case of the Navier-Stokes equation). It is
natural to conjecture that the power-law decay is a consequence of the
loss of smoothness of the solution during collapse; however we have
not been able to compute the exponent $\alpha$ analytically.

\begin{figure}[h]
  \centering
  \includegraphics[width=0.99\columnwidth]{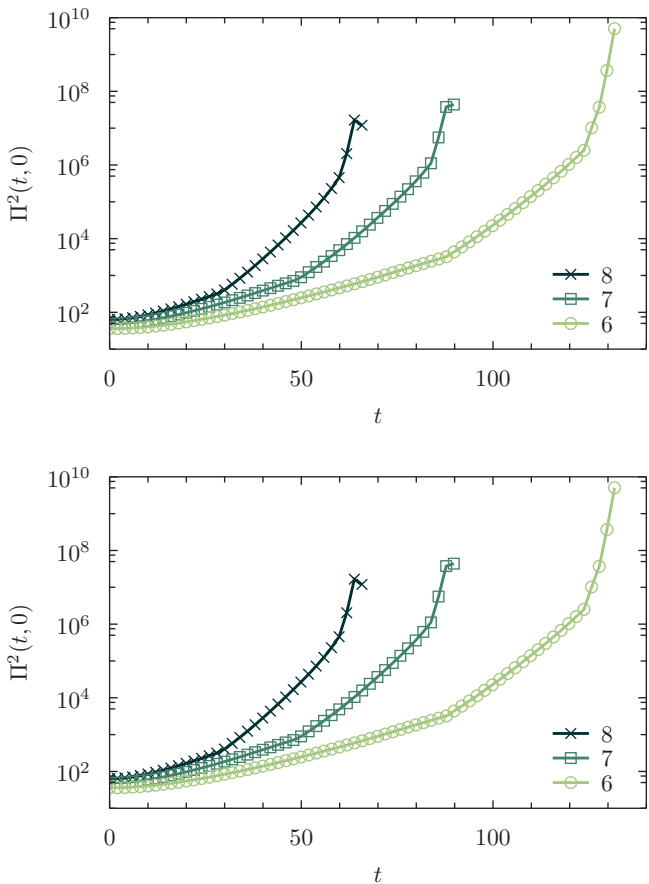}
  \vspace{1ex}

  \centering
  \includegraphics[width=0.99\columnwidth]{familyA_ricci_noscaled_neu.eps}
  \caption{ The analogue of Fig.~\ref{fig:ricci} for the Neumann
    boundary condition [for the same initial data \eqref{eq:idg}]. The
    time of apparent horizon formation exhibits the same type of
    scaling $t_{AH}\sim\varepsilon^{-2}$.}
  \label{fig:ricci_neu}
\end{figure}

Close parallels between the results presented in this work and
\cite{br, jrb} indicate that the turbulent behavior is not an
exclusive domain of asymptotically AdS spacetimes but a typical
feature of 'confined' Einstein's gravity with reflecting boundary
conditions. This answers the question about the role of the negative
cosmological constant $\Lambda$ posed at the end of \cite{br}: the
only role of $\Lambda$ is to generate the timelike boundary at spatial
and null infinity.

Finally, let us point out that instead of the Dirichlet condition at
the boundary of the cavity $\Pi(t,R)=0$ one can alternatively consider
the Neumann boundary condition $\Phi(t,R)=0$ (for which the energy
flux through the boundary vanishes as well). In the Neumann case the
eigenfrequencies of the linearized problem, determined by the
condition $\tan R\omega_j = R\omega_j$, are only asymptotically
resonant
\begin{equation*}
  \omega_{j}=\frac{\pi}{R} \left( j+\frac{1}{2} \right) +\mathcal{O}\left(j^{-1}\right),
  \quad j\to\infty\,, \quad j\in\mathbb{N}\,
  \label{eigen_neu}
\end{equation*}
nonetheless numerical simulations show similar turbulent behavior as
in the Dirichlet case (see Fig.~\ref{fig:ricci_neu}). This indicates that
the spectrum of linearized perturbations need not be fully resonant
for triggering the instability (cf. Ref. \cite{dhms} for an opposite
conclusion based on the nonlinear perturbation analysis).

I thank Piotr Bizo\'n and Andrzej Rostworowski for suggesting the
problem, encouragement, and discussions. This work was supported in
part by the NCN Grant No. NN202 030740 and Dean's Grant
No. DSC/000703. Most computations were performed on the supercomputer
Deszno at the Institute of Physics \linebreak of the Jagiellonian University.

\end{document}